\documentclass[a4paper,10.5pt,twocolumn,twoside]{article}

\usepackage{graphicx}
\usepackage{color}
\usepackage[a4paper, left=1.4cm, right=1cm, top=1.7cm, bottom=1cm]{geometry}
\usepackage[super,comma,sort&compress,square]{natbib}
\usepackage{amsmath}
\usepackage{float}
\usepackage{latexsym}
\usepackage{amssymb}

\usepackage{fancyhdr}
\usepackage[utf8x]{inputenc}
\usepackage[T1]{fontenc}

\begin{document}
\pagestyle{fancy}
\def\headrulewidth{0.5pt}
\def\footrulewidth{0pt}
\lhead{ACS Applied Materials \& Interfaces 9 (2017), 33250 -- 33256} 
\chead{}
\rhead{DOI: 10.1021/acsami.7b07665}

\lfoot{} 
\cfoot{}
\rfoot{}

\twocolumn[
  \begin{@twocolumnfalse}
  {\huge \bf Exchange Bias in the \boldmath$\lbrack$CoO/Co/Pd$\rbrack_{10}$ Antidot Large Area Arrays}

  \hspace{1.1cm}
  \parbox{.87\textwidth}{
    \vspace{4ex}
    \Large \textsf{Marcin Perzanowski,$^{1}$ Michal Krupinski,$^{1}$ Arkadiusz Zarzycki,$^{1}$  Andrzej Dziedzic,$^{2}$ Yevhen Zabila,$^{1}$ Marta Marszalek$^{1}$}
    \vspace{1ex} \\
    \normalsize $^{1}$ Institute of Nuclear Physics Polish Academy of Sciences, Deparment of Materials Science, Radzikowskiego 152, 31-342 Krakow, Poland  \\
    $^{2}$ University of Rzeszow, Center of Innovation \& Knowledge Transfer, Pigonia 1, 35-310 Rzeszow, Poland
    \vspace{1ex} \\
    \normalsize \text{email: Marcin.Perzanowski@ifj.edu.pl}

    \vspace{2ex} 
    \noindent
     \textbf{Abstract}: Magnetic nanostructures revealing exchange bias effect have gained a~lot of interest in recent years due to their possible applications in modern devices with various functionalities. 
In this paper, we present our studies on patterned [CoO/Co/Pd]$_{10}$ multilayer where ferromagnetic material is in a~form of clusters, instead of being a~continuous layer. 
The system was patterned using nanosphere lithography technique which resulted in creation of an assembly of well-ordered antidots or islands over a~large substrate area. 
We found that the overall hysteresis loop of the films consists of hard and soft components. 
The hard component hysteresis loop exhibits a~large exchange bias field up to $-11$~kOe. The patterning process causes a~slight increase of the exchange field as the antidot radius rises. 
We also found that the material on edges of the structures gives rise to a~soft unbiased magnetization component. 
     
     \vspace{2ex}
     DOI: 10.1021/acsami.7b07665
     
     \vspace{2ex}
     
    \vspace{3ex}
  }
  \end{@twocolumnfalse}
]

\section{Introduction}

Exchange bias effect occurs at the interface between a~ferromagnet (F) and an antiferromagnet (AF) after cooling in the magnetic field through AF N\'eel temperature, and is observed as a~magnetic hysteresis loop shift along the external field axis.\cite{Kiw01JMMM} 
Due to its interfacial origin the magnitude of this loop shift is inversely proportional to the thickness of the F layer.\cite{Nog99JMMM} 
The effect vanishes above the blocking temperature for exchange bias which, due to finite-size effects, is usually lower than AF N\'eel temperature. 
The exchange bias phenomenology has been described using the model of rough F-AF interface,\cite{Mal87PRB} domain state model,\cite{Mil00PRL} and by considering the role of uncompensated interfacial spins.\cite{Ge13N,Tak97PRL} 

In recent years materials exhibiting exchange bias effect have gained a~lot of attention due to their technological applications in the magnetic read heads,\cite{Par03IEEE} sensors and biosensors,\cite{Neg09APL,EhrS15} spintronic devices,\cite{Sue05APL,Ngu11APL,Gas13APL,Pol14APL} and as drug carriers.\cite{Iss13IJMS} 
The development of these technologies requires constant miniaturizing of the devices. 
One of the ways to achieve this goal is through the patterning of flat films. 
Besides the technological aspect, the studies of magnetic patterned structures are of scientific interest due to the fact that material properties can change depending on the geometry of a~specimen. 
There are several reports on exchange bias effect in various types of shapes and arrangements, such as rings,\cite{Tri09N} disks,\cite{Gil15N} arrays of antidots,\cite{Tri08APL} or dots,\cite{Car14JAP,Suc09APL} and core-shell nanostuctures.\cite{Sal16AMI,Shi14N}  

In this work, we consider an influence of a~patterning process on magnetic properties of exchange biased multilayer. 
First, we raise an issue of decreasing ferromagnet's thickness below the limit for continuous layer formation. 
Previous experiment works were focused on systems with the continuous F layers having thicknesses in the nanometer range and for which the influence of thermal fluctuation related to the appearance of the superparamagnetic behavior was not an important factor. 
Second, we examine the changes in magnetic properties induced by a~patterning process. 
Different technological applications require particular types of structure, and in attempt to tackle this subject we study an evolution of magnetic properties starting from a~flat multilayer and going through systems with antidots to an assembly of separated islands.

For our studies we chose the AF/F CoO/Co interface which is being considered a~model structure for investigations of exchange bias effect.\cite{Men14JAP,Dob12PRB,Dia14JAP} 
The Pd layers were introduced in the structure to obtain the perpendicular magnetic anisotropy associated with the Co-Pd system.\cite{Car85APL,Car03APL} 
This property is particularly important from the point of view of data storage and memory devices. 
Our previous research showed\cite{Per16AMI} that the reduction of the ferromagnet thickness below the limit for continuous layer formation leads to the creation of Co-Pd clusters which are being progressively blocked with the decreasing temperature. 
This behavior resulted in a~large enhancement of the exchange bias field up to $-11$~kOe. 
In this paper we concentrate on the role of the patterning by nanosphere lithography on the magnetic properties of exchange biased system.

\section{Experimental section}

The arrays of antidots and separated islands were prepared using the nanoshpere litography technique.\cite{Kos04NL,Kru15N,Kru17N} 
Prior to the [CoO/Co/Pd]$_{10}$ multilayer deposition flat Si(100) substrates were covered by an assembly of self-organized polystyrene (PS) spherical nanoparticles with average diameter of 784~nm. 
Next, the PS nanoparticles were etched in RF-plasma to decrease the size of the spheres while their position remained unchanged. 
The etching process was carried out for 800~s, 400~s, and 175~s in oxygen and argon atmosphere.

The system was deposited simultaneously on five different substrates: three with plasma-etched PS nanoparticles, one with unprocessed nanoparticles, and the flat system as a~reference. 
The deposition process was carried out using thermal evaporation at room temperature in the ultrahigh vacuum chamber under the pressure of $10^{-7}$~Pa. 
First, the substrates were covered by 5-nm-thick Pd buffer layer. 
Next, the $0.5$~nm of Co was deposited and oxidized for 10 minutes in the oxygen atmosphere at the pressure of $3 \times 10^{2}$~Pa. 
Then, the oxide layer was covered by $0.3$~nm of Co and $0.9$~nm of Pd. 
After 10 repetitions of CoO/Co/Pd trilayer a~2-nm-thick Pd layer was deposited as the capping protective coating. 
After the deposition the PS nanoparticles were removed by ultrasonic cleaning in toluene. 
This resulted in the creation of the arrays of antidots, in case of etched PS nanospheres, or separated islands for the situation when the PS nanoparticles were not plasma-treated.

The Transmission Electron Microscopy (TEM) investigations were carried out using FEI Tecnai Osiris device with 200~keV primary electron beam. 
The Scanning Electron Microscopy (SEM) images were obtained using FEI Versa 3D machine operated at 5~keV of the primary electron beam. 
Magnetic Zero-Field-Cooled (ZFC), Field-Cooled (FC), and hysteresis loops measurements were done using a~Quantum Design MPMS XL SQUID magnetometer. 
Prior to the ZFC measurement a~demagnetized sample was cooled down to 5~K without the external magnetic field. 
The ZFC magnetization curves were recorded during sample heating up to 300~K in a~magnetic field of 500~Oe. 
The FC curves were obtained during sample cooling from 300~K to 5~K in the same magnetic field as for ZFC measurements. 
The hysteresis loops were measured at 10~K after cooling a~sample in the field of $+50$~kOe. 
All magnetic measurements were carried out in out-of-plane (OOP) geometry with the external magnetic field perpendicular to the substrate plane. 
The data were corrected for the diamagnetic background from the sample holder.

\section{Results and discussion}

\begin{figure*}[ht!]
 \centering
 \includegraphics[height=4.75cm]{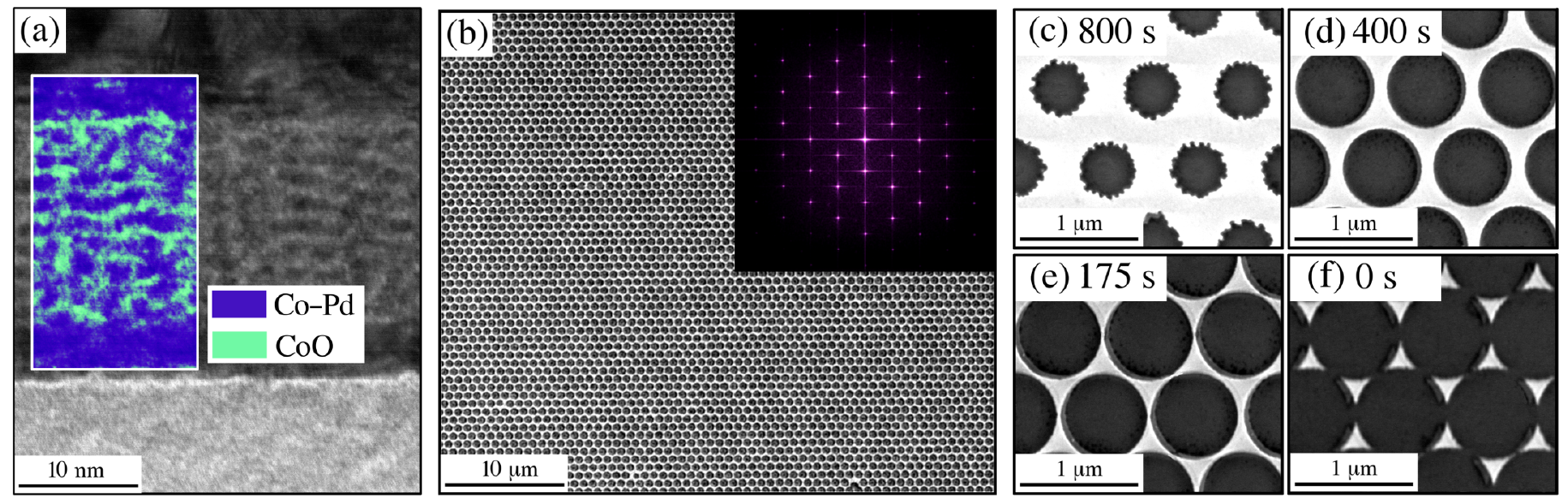}
 \caption{(a) TEM cross-section image of the [CoO/Co/Pd]$_{10}$ flat multilayer. 
 In the false-colored inset the Co-Pd and CoO regions are indicated. 
 (b) Large-area SEM image of the [CoO/Co/Pd]$_{10}$ multilayer deposited on PS nanoparticles etched for 175~s. 
 The inset shows the Fourier transform of the initial SEM picture. 
 (c) -- (f) SEM images taken for multilayer deposited on PS nanoparticles etched for 800~s, 400~s, 175~s, and without PS etching. 
 The bright fields are magnetic multilayer, the dark regions are silicon substrate.}
 \label{Fig_1}
\end{figure*}

The TEM cross-section image of the flat reference sample is presented in Figure~\ref{Fig_1}a. 
The total thickness of the [CoO/Co/Pd]$_{10}$ multilayer is 28~nm and, due to the oxidation process, is larger than the total thickness of the deposited material. 
It can be seen that the ingredient layers are not continuous due to the high interface jaggedness. 
The thickness of the deposited non-oxidized Co is only $0.3$~nm, which led to the Co atoms intermixing with the adjacent Pd layers. 
Because of this, in the TEM image it is only possible to distinguish between CoO and Co-Pd regions. 
Taking into account the jagged structure of the interfaces as well as the intermixing between the materials it is  clear that the system has a~nanogranular structure with ferromagnetic Co-Pd clusters and antiferromagnetic CoO grains. 

The patterning procedure resulted in the creation of 2D hexagonal lattices of antidots or islands. 
Figure~\ref{Fig_1}b shows a~representative large-area SEM image of the multilayer on PS nanoparticles etched for 175~s. 
The data and its Fourier transform show that the antidot arrangement is highly ordered and lattice defects are rare. 
The SEM close-ups of the patterned samples are shown in Figure~\ref{Fig_1}c -- f. 
The smallest antidots (dark fields in the image) with the average radius of $r_{\mathrm{a}}=217$~nm are fabricated after 800~s of plasma etching. 
The average antidot radius increased to 343~nm and 367~nm, after 400~s and 175~s of plasma etching, respectively. 
In the case of unprocessed PS nanoparticles the multilayer forms an assembly of separated triangular-like-shaped islands. 
The geometry of these islands arrangement corresponds to the average antidot radius of 395~nm.

\begin{figure*}[t]
 \centering
 \includegraphics[height=5cm]{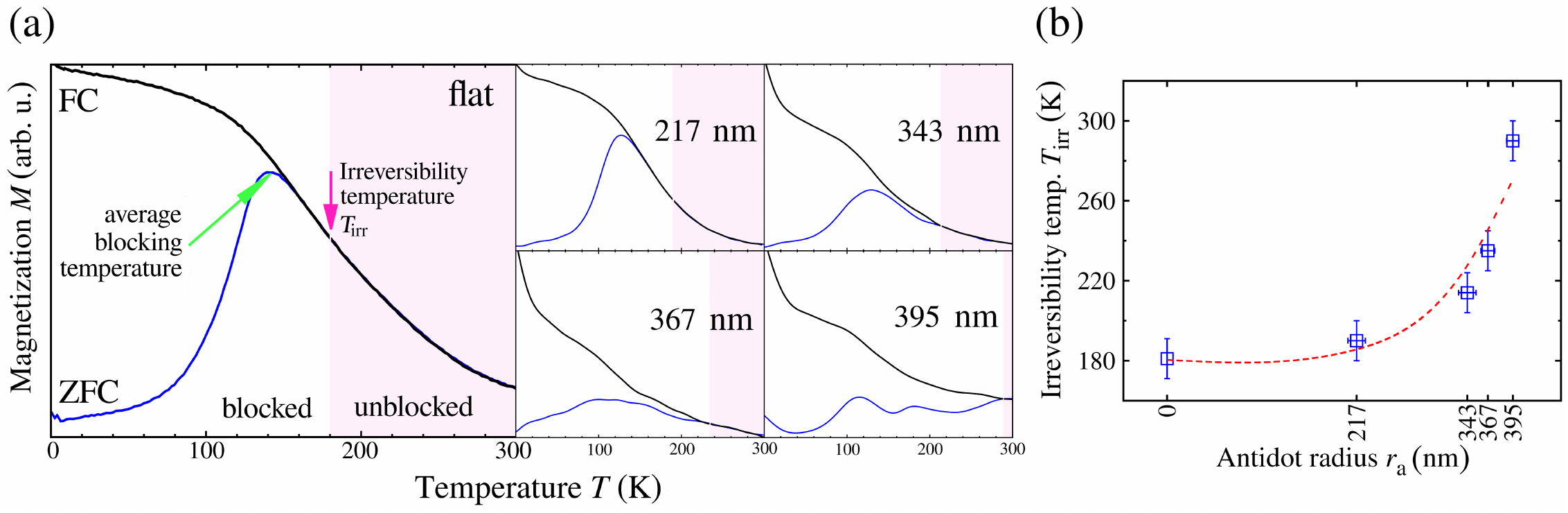}
 \caption{(a) Zero-Field-Cooled (ZFC) and Field-Cooled (FC) magnetization curves measured for the [CoO/Co/Pd]$_{10}$ multilayer deposited on the flat and patterned substrates. 
 (b) The irreversibility temperature $T_{\mathrm{irr}}$ dependence on the antidot radius $r_{\mathrm{a}}$. 
 The dashed line is a~guide for the eye.}
 \label{Fig_2}
\end{figure*}

The ZFC and FC magnetization curves for all samples are shown in Figure~\ref{Fig_2}a. 
In the case of the flat system the irreversibility temperature $T_{\mathrm{irr}}$, above which the ZFC and FC curves overlap, is approximately 180~K. 
The overlap means that above this temperature the anisotropy energy of the Co-Pd clusters is too weak to overcome thermal fluctuation energy and the system reveals superparamagnetic behavior.\cite{Bed09JoP} 
Below this temperature the low thermal energy allows the Co-Pd superspins blocking. 
The blocked Co-Pd clusters can be exchange-coupled with antiferromagnetic CoO grains inducing exchange anisotropy in the system, as demonstrated in our previous paper.\cite{Per16AMI} 
Additionally, we found that below $T_{\mathrm{irr}}$ the interactions between blocked Co-Pd clusters have superferromagnetic nature. 
Qualitatively similar behavior is observed in the ZFC/FC curves for the patterned samples. 
However, as the antidot radius increases, the irreversibility temperature increases, reaching the value close to the room temperature for the largest antidot size (Figure~\ref{Fig_2}b). 
In each ZFC curve for antidots the maximal magnetization, corresponding to the average blocking temperature of the Co-Pd clusters, is between 120~K and 145~K. 
The results also show that for larger antidot radius the breadth of this maximum increases. 
For the sample with separated islands the ZFC maximum splits up into two separated signals observed at approximately 115~K and 180~K. 

\begin{figure*}[t]
 \centering
 \includegraphics[height=8cm]{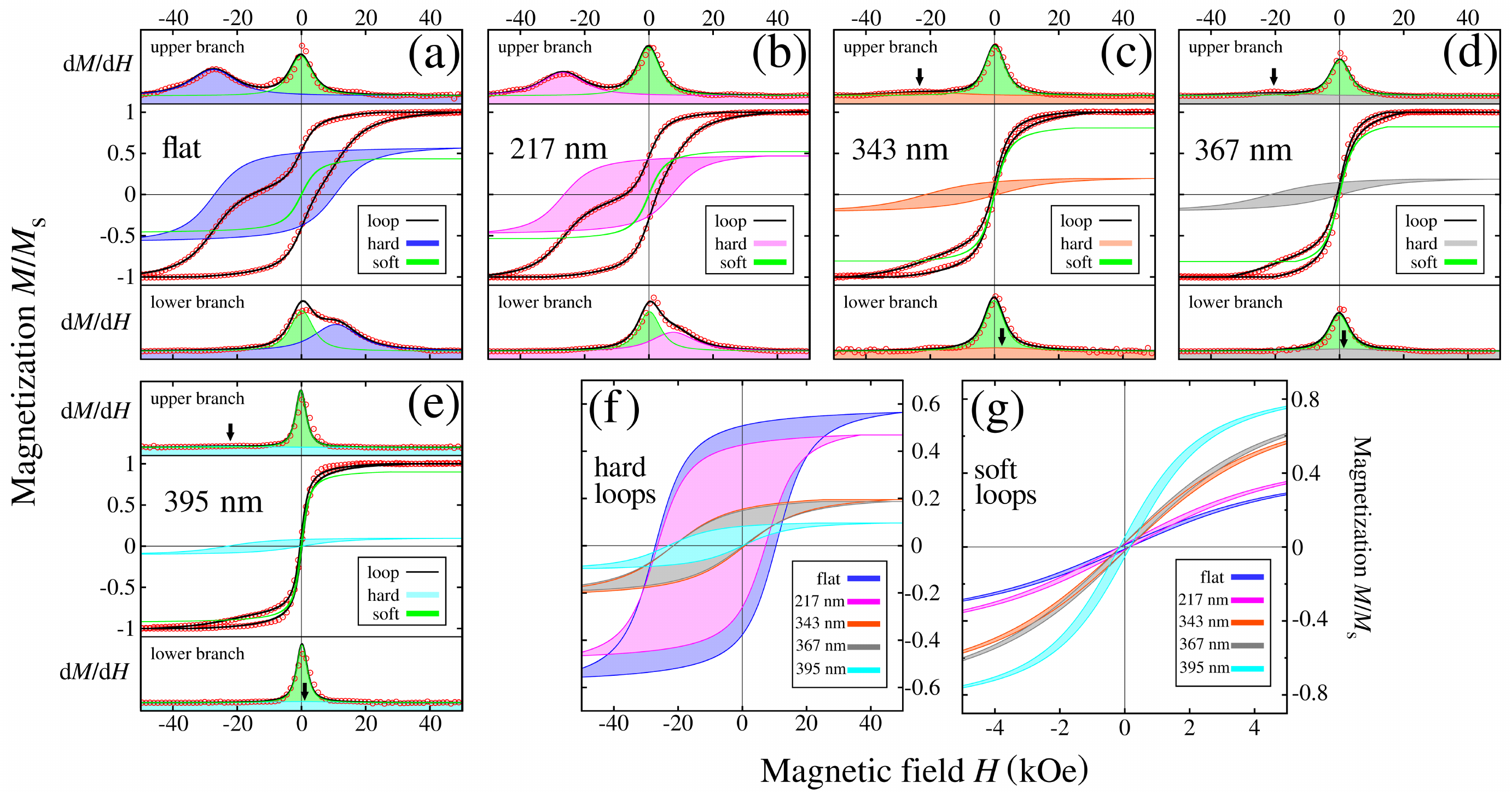}
 \caption{(a) -- (e) Hysteresis loops measured at 10~K after field cooling in $+50$~kOe. 
 The panels above and below each loop show first derivatives of the upper and lower branches of the magnetization curve. 
 Red points are the experimental data. Black solid lines are fits of the overall loops (see text). 
 The hard and soft loops are marked with colored fields. 
 In figures (c), (d), and (e) the positions of the hard component d$M$/d$H$ maxima are marked with the black arrows. 
 The antidot radii are indicated in all figures. 
 Comparisons of the fitted hard and soft hysteresis loops are shown in Figures (f) and (g), respectively.}
 \label{Fig_3}
\end{figure*}

The hysteresis loops measured after field cooling in $+50$~kOe from room temperature to 10~K for all [CoO/Co/Pd]$_{10}$ systems are shown in Figure~\ref{Fig_3}. 
It shows that the loop for the flat film (Figure~\ref{Fig_3}a) has an asymmetric shape. 
The first derivatives d$M$/d$H$ of the upper and lower magnetization branches have two separated maxima corresponding to different switching fields, and showing that the system consists of two magnetic phases. 
The high-field switching fields (blue marks in Figure~\ref{Fig_3}a), further denoted as a~hard component, have their maxima for several kOe with a~clear center shift towards the negative external field. 
This effect confirmes that the blocked Co-Pd clusters are coupled to the antiferromagnetic CoO grains introducing the exchange anisotropy to the system and leading to the exchange bias effect. 
The exchange anisotropy field of $-11$~kOe observed in the loop is a~few times larger than for Co/CoO bilayers with continuous ferromagnetic layer having thickness in the nanometer range.\cite{Dob12PRB,Dia14JAP} 
In the case of the continuous layers the decrease of the F thickness leads to the rise of exchange bias anisotropy field.\cite{Nog99JMMM}. 
However, in the investigated system the Co thickness is in a~sub-nanometer range and this relation is no longer valid. 
We observe fragmentation of the layers and the system is in a~form of small AF and F objects interacting with each other, which can be considered a~case similar to nanocomposites. 
Magnetic Co/Pd/CoO composites revealing the exchange bias effect have not been studied yet, but the structure of our system can be compared to an assembly of core-shell nanoparticles.   
Baaziz et al.\cite{Baa13JPC} found a~remarkably large exchange bias field for the core-shell nanoparticles.   
This effect can be explained by the increased number of uncompensated spins present at the interfaces in such systems, as proposed by Iglesias et al.\cite{Igl07JoP} and Vasilakaki et al.\cite{Vas15SR}
Therefore, it can be expected that the increased number of uncompensated spins at the AF-F interfaces arising from the nanogranular structure of our systems is responsible for the large exchange anisotropy field observed in the results. 

The low-field d$M$/d$H$ maxima (green marks in Figure~\ref{Fig_3}a) correspond to a~soft component of magnetization and are centered around zero external field. 
Both switching fields in this case have an absolute value of approximately 200~Oe. 
Since the constituent layers are not continuous and the system has a~nanogranular microstructure it can be expected that, besides the blocked Co-Pd clusters coupled to the CoO grains, there are also ferromagnetic parts of the layer not accompanied by antiferromagnetic cobalt oxide. 
Since they do not contribute to the AF/F exchange anisotropy their influence on the hysteresis loop shape is seen as a~soft component loop. 
A~similar asymmetric loop shape with two switching fields in each branch can also be observed for the patterned systems (Figure~\ref{Fig_3}b -- e), demonstrating that in all cases the overall magnetization curve consists of two components with similar origins.

\begin{figure}[t]
 \centering
 \includegraphics[height=6.5cm]{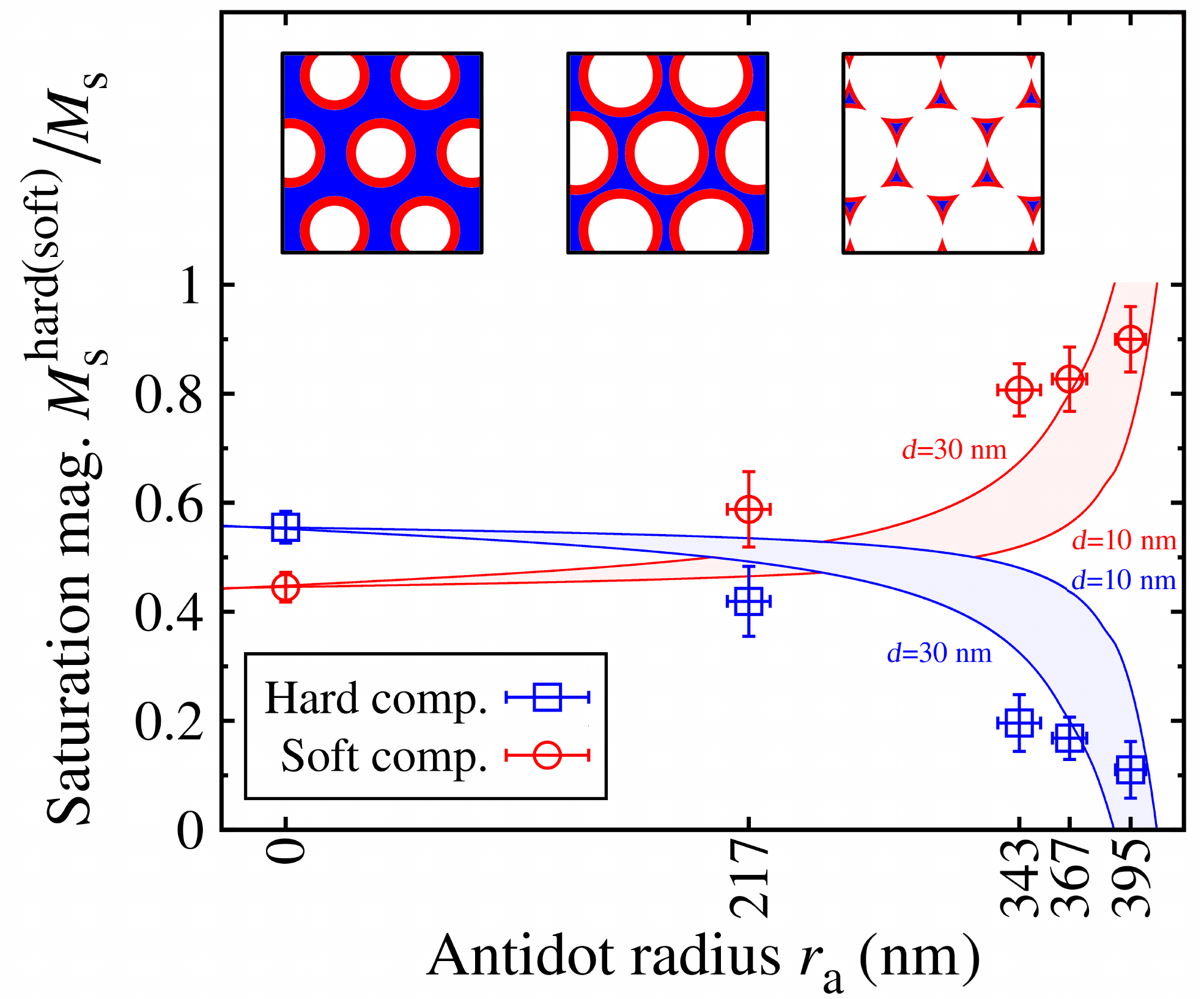}
 \caption{The dependencies of the saturation magnetizations of the hard $M_{\mathrm{s}}^{\mathrm{hard}}$ and soft $M_{\mathrm{s}}^{\mathrm{hard}}$ components on the antidot radius~$r_{\mathrm{a}}$. 
 The values were obtained from the fits presented in Figure~\ref{Fig_3} and are normalized to the saturation magnetization of the overall loop $M_{\mathrm{s}}$. 
 The red and blue fields mark the dependencies of the saturation magnetization on the antidot radius for different thicknesses $d$ of the edge region. 
 The inset shows changes of the inner volume (blue) and edge regions (red) for increasing antidot radius.}
 \label{Fig_4}
\end{figure}

The quantitative information about coercivities of the hard $H_{\mathrm{c}}^{\mathrm{hard}}$ and soft $H_{\mathrm{c}}^{\mathrm{soft}}$ components  and their saturation magnetizations $M_{\mathrm{s}}^{\mathrm{hard}}$ and $M_{\mathrm{s}}^{\mathrm{soft}}$ were obtained by fitting each hysteresis branch using the following expression:\cite{Ste94JAP,Kuz99JMMM}
\begin{eqnarray*}
 M(H)  =  \frac{2}{\pi}  M_{\mathrm{s}}^{\mathrm{hard}} \mathrm{arctan} \left(g^{\mathrm{hard}} \left[ \frac{H-H_{\mathrm{c}}^{\mathrm{hard}}}{H_{\mathrm{c}}^{\mathrm{hard}}} \right] \right) + \\ \frac{2}{\pi} M_{\mathrm{s}}^{\mathrm{soft}} \mathrm{arctan} \left(g^{\mathrm{soft}} \left[ \frac{H-H_{\mathrm{c}}^{\mathrm{soft}}}{H_{\mathrm{c}}^{\mathrm{soft}}} \right] \right)  , \ \ \ \ 
\end{eqnarray*}
where $g^{\mathrm{hard(soft)}}$ represents a~slope of a~magnetization curve. 
The fitted hysteresis loops are shown in Figures~\ref{Fig_3}a -- e where the green loops represent the unbiased soft component and the colored loops correspond to the hard biased component.
The comparisons of the fitted hard and soft hysteresis loops are presented in Figures~\ref{Fig_3}f -- g, respectively.
The dependencies of the saturation magnetizations on the antidot radius $r_{\mathrm{a}}$ are given in Figure~\ref{Fig_4}. 
The figure shows that as the antidot radius increases the soft component contribution rises as well, unlike the hard component which contribution decreases inversely proportional.  
The shape of the $M_{\mathrm{s}}^{\mathrm{soft(hard)}} \left(r_{\mathrm{a}}\right)$ dependencies indicate the presence of the edge effects which affects the magnetic properties of the system. 
Since the multilayer was evaporated through the mask of the PS nanospheres, one can anticipate the occurrence of the shadowing during the deposition process. 
Therefore, it can be expected that the material on the edges has altered magnetic properties, with respect to the inner multilayer.\cite{Kru17N,Sha08PRB,Lee11APL} 

The fits presented in Figure~\ref{Fig_3}g revealed that for each system the soft component has coercivity of 200 -- 300~Oe. 
In the case of the flat film the influence of the edge effects can be excluded and the observed soft component arises from the inner Co-Pd clusters. 
These clusters are not coupled to the antiferromagnetic grains and their presence is caused by the high interface jaggedness. 
For the patterned systems, the occurrence of the soft component has two sources --- the inner Co-Pd clusters, as for the flat samples, and the uncoupled clusters placed at the structure edges formed due the patterning process. 
It is worth noting that the inner uncoupled Co-Pd clusters reveal the same coercivity as these placed on the structure edges.

The saturation magnetization of the hard $M_{\mathrm{s}}^{\mathrm{hard}}$ and soft $M_{\mathrm{s}}^{\mathrm{soft}}$ components is proportional to the volumes $V_{\mathrm{hard}}$ and $V_{\mathrm{soft}}$ occupied by the corresponding magnetic phases. 
After normalization to the saturation magnetization of the overall loop $M_{\mathrm{s}}$ the following magnetizations can be expressed as $(V_{\mathrm{hard(soft)}})/(V_{\mathrm{hard}}+V_{\mathrm{soft}})$. 
In the case of the flat film, due to the lack of the structure edges, the overall sample amount equals to the inner volume $V_{\mathrm{inn}}$. 
In this case the hard component contribution $M_{\mathrm{s}}^{\mathrm{hard}}$ to the overall loop saturation $M_{\mathrm{s}}$ is $0.55$ which means that such volume fraction of the ferromagnetic material is exchange coupled to the antiferromagnetic grains. 
The remaining 45\% of the ferromagnetic material volume gives rise to the soft magnetization component $M_{\mathrm{s}}^{\mathrm{soft}}$. 
It can be assumed that this relation is fulfilled also for the patterned samples regarding their inner volume (blue areas, inset in Figure~\ref{Fig_4}). 
Moreover, for the patterned films the soft component also comes from the material volume located at the altered edges $V_{\mathrm{edge}}$ (red areas, inset in Figure~\ref{Fig_4}). 
Therefore, the volume of the hard component is $0.55 V_{\mathrm{inn}}$, and the volume of the soft component is $0.45 V_{\mathrm{inn}} + V_{\mathrm{edge}}$. 
For the constant period of the structure pattern both $V_{\mathrm{inn}}$ and $V_{\mathrm{edge}}$ are dependent on the antidot radius $r_{\mathrm{a}}$ and the edge thickness $d$, and the expressions for the saturation magnetization dependencies can be written as:
\[
M_{\mathrm{s}}^{\mathrm{hard}} (r_{\mathrm{a}},d) \propto \frac{0.55 V_{\mathrm{inn}} (r_{\mathrm{a}},d)}{V_{\mathrm{inn}} (r_{\mathrm{a}},d)+V_{\mathrm{edge}} (r_{\mathrm{a}},d)}
\]
and
\[
M_{\mathrm{s}}^{\mathrm{soft}} (r_{\mathrm{a}},d) \propto \frac{0.45 V_{\mathrm{inn}} (r_{\mathrm{a}},d) + V_{\mathrm{edge}} (r_{\mathrm{a}},d)}{V_{\mathrm{inn}} (r_{\mathrm{a}},d) + V_{\mathrm{edge}} (r_{\mathrm{a}},d)} . 
\]
As the antidot radius increases, the edge volume $V_{\mathrm{edge}}$ becomes dominant over the inner volume $V_{\mathrm{inn}}$. 
This behavior is responsible for the continuous rise of the soft component contribution observed in our results.

Taking into account the hexagonal arrangement of the pattern, the application of the following model to our results reveals that the altered structure edge thickness $d$ is between 10~nm and 30~nm (see blue and red curves in Figure~\ref{Fig_4}).  
This value is larger than the edge thickness reported by Shaw et al.\cite{Sha08PRB} most likely due to the different patterning procedure.  
On the other hand, Krupinski et al.\cite{Kru17N} applied the same patterning procedure and observed the edge thickness of 12~nm. 
However, in this case the Co layers were not oxidized. 
The oxidation, used in this paper, can induce materials intermixing and, together with the shadowing effect of the PS mask, can modify the thickness of the edges. 
The thickness of the nanostructure edge, similar to our results, was reported by Cast\'an-Guerrero et al.\cite{Cas14PRB} for the magnetic antidots made using the ion beam technique. 
Lee et al.\cite{Lee11APL} studied magnetic nanodots with a diameter ranging from 16~nm to 44~nm produced using nanoimprint lithography, and estimated the damage edge region for 40\% of the nanostructure size. 
The influence of the damaged nanostructure edges was also reported by Rahman et al.\cite{Rah07APL,Rah10PRB} for the magnetic multilayers deposited on anodized alumina oxides; however, no precise determination of the edge thickness was given in these works.

The gradual change of the soft and hard component contributions is also reflected in the ZFC curves (Figure~\ref{Fig_2}a). 
The figure shows that for the rising antidot radius the blocking temperature distribution of the Co-Pd clusters becomes broader while the temperature at which it reaches the maximum remains unchanged. 
The ZFC curve in our case can be considered as a~convolution of the two contributions corresponding to the hard and soft magnetization components. 
Therefore, for increasing antidot radius the blocking temperature distribution of the soft component should become better pronounced while the hard component contribution has to turn weaker. 
The gradual ZFC curve broadening is observed especially from the higher temperature side which implies that the soft component has higher average blocking temperature than the hard component. 
It is particularly well visible in the ZFC curve for the system with islands ($r_{\mathrm{a}}=395$~nm) where two separated maxima are observed. 
This rising contribution of the soft component having higher blocking temperature than the other one causes the observed shift of the irreversibility temperature towards room temperature as the antidot radius increases (Figure~\ref{Fig_2}b).

The hard component hysteresis loops, extracted from the overall loops, are shown in Figure~\ref{Fig_3}f. 
The ratio between the magnetic remanence and the saturation magnetization for flat sample and for system with 217~nm antidots is approximately 0.9 which demonstrates that in these cases the easy axis of magnetization of the hard component points in the direction perpendicular to the system plane. 
Such magnetization bearing is caused by the large out-of-plane anisotropy revealed by the Co-Pd material.\cite{Car85APL,Car03APL} 
For the remaining systems the ratio between the remanence and saturation magnetizations is in range from 0.5 to 0.6 suggesting rather isotropic distribution of magnetization axes orientation. 
It also can be seen that in all cases the loops have symmetric shape, with respect to the certain shift along the field axis. 
This means that for both ascending and descending branches of the loop the magnetization reversal process takes place in the same way. 
Similar symmetry of the hysteresis loop shape was reported by Baltz et al.\cite{Bal05PRL} for sub-100~nm permalloy/IrMn nanostructures. 
On the other hand, Tripathy et al. argued that for the Ni$_{80}$Fe$_{20}$/Ir$_{75}$Mn$_{25}$ antidot arrays the reversal mechanism for the upper and lower branch is different.\cite{Tri08APL} 
However, contrary to our results, in both these studies the ferromagnetic and antiferromagnetic layers were relatively thin and continuous. 

\begin{figure*}[t]
 \centering
 \includegraphics[height=9.1cm]{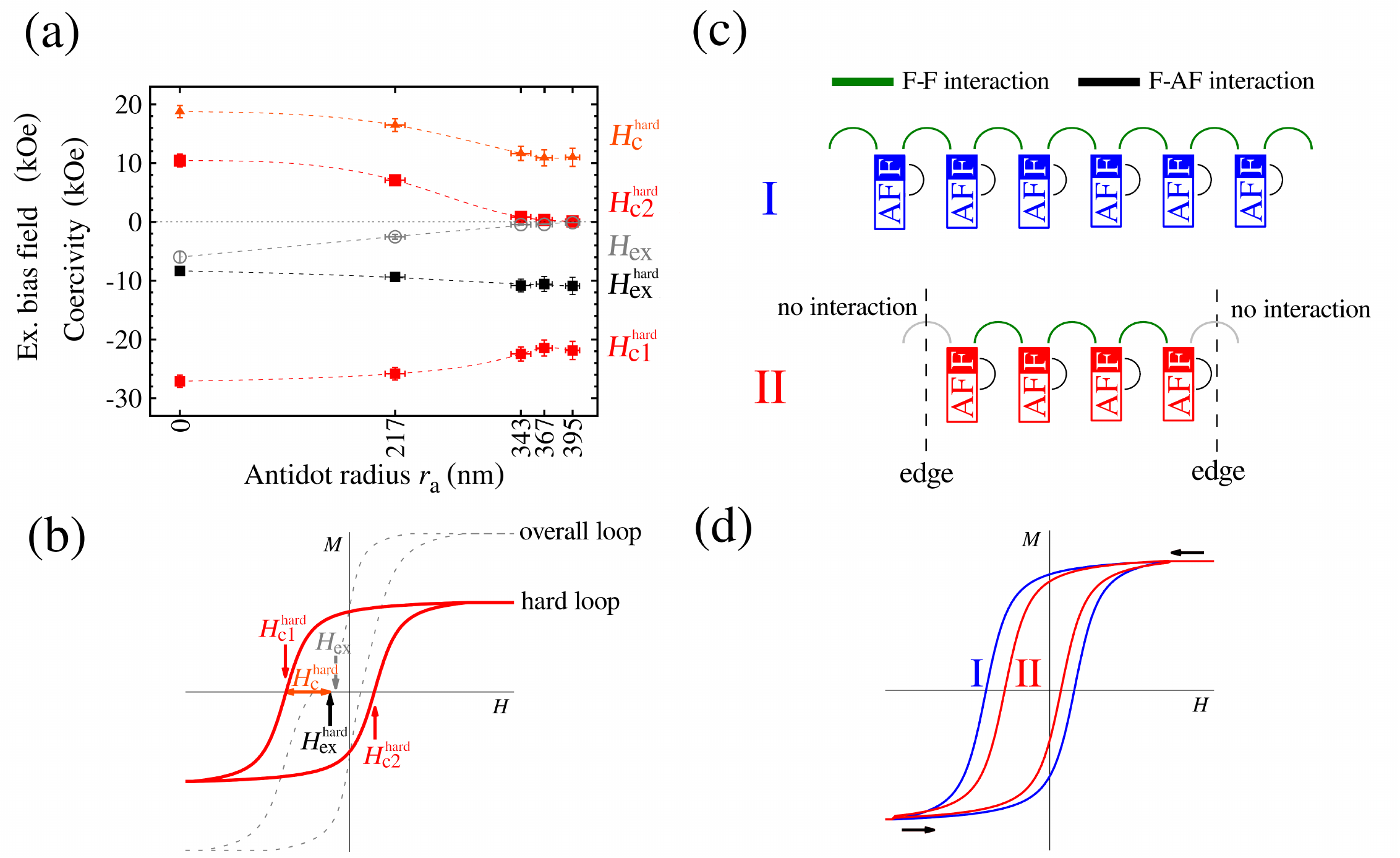}
 \caption{(a) Dependencies of coercivity and the exchange bias field on the antidot radius measured for [CoO/Co/Pd]$_{10}$ samples. 
 (b) A~representative hysteresis loop for the patterned exchange biased system with marks of the quantities shown in Fig.~a. 
 (c) Schematic representation of interactions between ferromagnetic clusters (F) and antiferromagnetic grains (AF) in case of: (I) quasi-infinite flat sample, (II) patterned system with edges. 
 (d) A~representative hysteresis loops corresponding to the cases (I) and (II) from Fig.~c.}
 \label{Fig_5}
\end{figure*}

The dependencies of coercivity and the exchange bias field on the antidot radius measured for our CoO/Co/Pd systems are presented in Figure~\ref{Fig_5}a. 
The values were obtained from the hard component magnetization curves shown in Figure~\ref{Fig_3}f. 
The hysteresis loop parameters under investigation given in Figure~\ref{Fig_5}a are schematically marked in Figure~\ref{Fig_5}b. 
We show that the increase of the antidot radius $r_{\mathrm{a}}$ results in the decrease of the absolute value of both $H_{\mathrm{c1}}^{\mathrm{hard}}$ and $H_{\mathrm{c2}}^{\mathrm{hard}}$ coercive fields corresponding to the upper and lower branches of the magnetization curve, which leads to the reduction of the loop coercivity~$H_{\mathrm{c}}^{\mathrm{hard}}$. 
The exchange bias field of the overall loop $H_{\mathrm{ex}}$ is being constantly reduced as the antidot radius increases, reaching almost zero for the system with separated islands. 
This decrease is associated with the constant growth of the unbiased soft component contribution as the antidots radius becomes larger (see Figure~\ref{Fig_4}). 
The absolute value of the exchange bias field of the hard component $H_{\mathrm{ex}}^{\mathrm{hard}}$ gradually rises from $8.5$~kOe for the flat sample to $11$~kOe for the system with separated islands.

Let us now consider a~quasi-infinite flat thin film where influence of the specimen edges is negligible, and material is in form of ferromagnetic clusters (F) connected to antiferromagnetic grains (AF). 
For the sake of clarity we restrict our line of reasoning to the single AF-F CoO-Co/Pd block in the mutlilayer structure. 
In such case both the F and AF materials are placed on top of another, as it is shown in the panel (I) of Figure~\ref{Fig_5}c. 
First type of magnetic interaction in this case is a~ferromagnet-ferromagnet F-F interaction between neighboring clusters which takes place along the direction parallel to the film plane. 
Since the material under consideration is quasi-infinite each F cluster is surrounded by other F clusters and is magnetically coupled to them. 
This F-F interaction gives rise to the coercivity of the system. 
Second type of magnetic interaction in this case is a~ferromagnet-antiferromagnet F-AF exchange coupling responsible for the exchange bias effect appearance and leading to a~hysteresis loop shift from the center position. 

The following argumentation is valid for the case with no vertical interaction between the F clusters placed in the subsequent layers in the actual multilayer system, throughout the AF grains. 
Assuming that the presence of the AF material is a~sufficient barrier for such vertical F-F interaction to occurr, this model situation can be illustrated in our results by the flat [CoO/Co/Pd]$_{10}$ film. 
A~representative hysteresis loop for the case of the quasi-infinite flat sample is given in blue in Figure~\ref{Fig_5}d. 
It is worth noting that in this system, due to the interface jaggedness, there are Co-Pd clusters connected along the direction perpendicular to the sample plane without any CoO grain between them (see Figure~\ref{Fig_1}a). 
The presence of these objects is not included in our model. 
However, they do contribute only to the soft component magnetization loop and do not have influence on the hard loop coercivity or the exchange bias field. 

A~different situation takes place when the specimen is confined by edges perpendicular to the sample plane (see panel (II) in Figure~\ref{Fig_5}c). 
In such case the F clusters placed on these edges experience lack of the F-F interaction from the exposed side. 
Therefore, the coupling strength experienced by the edge F clusters is weaker than by the objects located inside the specimen. 
In this situation it can be expected that such edge clusters reverse their magnetization direction under lower external magnetic field than the inner objects which have larger number of neighbouring Co-Pd clusters to interact with. 
In this way, the coercive fields of the upper branch and lower branches of the hysteresis loop measured for the patterned samples with the edges, have to be lower than in case of the quasi-infinite flat system (see the red loop in Figure~\ref{Fig_5}d). 
The following reduction of coercivity is observed in our results. 
Moreover, this effect is more pronounced for samples having larger antidot radius (Figure~\ref{Fig_5}a) which is related to the increasing ratio between the number of the F clusters placed on the structure edges and the number of the inner F clusters as the antidot size grows. 

Contrary to the F-F interaction, the F-AF coupling takes place along the direction perpendicular to the film plane (Figure~\ref{Fig_5}c). 
Therefore, the magnitude of the exchange coupling interaction should not change due to the presence of the specimen edges and result in lack of influence of the system lateral geometry on the exchange bias field. 
However, in our results we observe a~slight rise of the exchange bias field $H_{\mathrm{ex}}^{\mathrm{hard}}$ for increasing antidot radius. 
This discrepancy between the presented model and the obtained $H_{\mathrm{ex}}^{\mathrm{hard}}$ values may be caused by a~change in the F clusters size distribution upon the patterning process, while the average cluster size remains constant as the ZFC/FC curves show. 
The other explanation is that the edge regions contain not only the soft unbiased material but there are also small Co-Pd clusters coupled to the AF grains which contribute to the exchange anisotropy. 
As the antidot radius increases and the input from the edges becomes dominant, it also slightly enhances the loop shift. 

\section{Conclusions}

In this paper, we have studied the magnetic properties of the exchange biased [CoO/Co/Pd]$_{10}$ antidot arrays. 
The ferromagnetic material in this system has been in form of Co-Pd clusters. 
It reveals superparamagnetism at higher temperature and being blocked when temperature drops down. 
We have observed that the major hysteresis loops for all systems consist of hard and soft components. 
The hard component exhibits large exchange bias field up to $-11$~kOe, which is several times larger than for systems with continuous ferromagnetic layer. 
Moreover, we have found that the patterning process induces a~slight rise of the bias field for increasing antidots radius. 
We have observed that the transition from the flat [CoO/Co/Pd]$_{10}$ multilayer throughout the pattern of antidots to the assembly of separated islands results in the gradual increase of the soft unbiased magnetization component coming from the material placed at the edges of the structures.

\bibliographystyle{plainnat}
\vspace{1.1ex}
\begin{center}
 $\star$ $\star$ $\star$
\end{center}

\vspace{-9ex}

\setlength{\bibsep}{0pt}
\renewcommand{\bibnumfmt}[1]{$^{#1}$}

\end{document}